\documentclass[reprint,prb,showpacs]{revtex4-1}

\usepackage{graphicx}
\usepackage{hyperref}
\usepackage{epstopdf}
\usepackage{caption}
\usepackage{subcaption}

\begin{document}

\title{Magnetic instability and $f-d$ hybridization in CeFe$_2$ on substituting Cr, Ag and Au for Fe}

\author{Rakesh Das}
\affiliation{Department of Physics, Indian Institute of Technology Kharagpur, Kharagpur 721302, India}
\affiliation{Department of Applied Science, Haldia Institute of Technology, Haldia 721657, India}
 \email{dasrakesh@phy.iitkgp.ernet.in/icon7117@gmail.com}
 
\author{Mukul Gupta}
\affiliation{UGC-DAE Consortium for Scientific Research, University Campus, Khandwa Road, Indore 452001, India}
 \email{mgupta@csr.res.in/dr.mukul.gupta@gmail.com}

\author{S. K. Srivastava}
\affiliation{Department of Physics, Indian Institute of Technology Kharagpur 721302, India}
 \email{sanjeev@phy.iitkgp.ernet.in}

\date{\today}

\begin{abstract}

Hybridization between Ce $f$ and conduction $d$ states has been speculatively known to be one of the mechanisms responsible for magnetic instability of the ferromagnetic ground state of CeFe$_2$. Substituting Fe by small amounts of certain elements stabilizes it to an antiferromagnetic state below the Curie temperature via a first-order second phase transition. In the present work, we seek any direct relation between the $f-d$ hybridization and the second transition by measuring primarily dc magnetization and Ce M$_{4,5}$ edge x-ray absorption spectra of Ce(Fe$_{1-x}$M$_x$)$_2$ pseudobinaries, with M = Cr, Ag and Au. X-ray diffraction and x-ray photoelectron spectroscopy measurements are also performed essentially to monitor the quality of the samples. Whereas Cr impurity is found to cause the second transition, Ag and Au apparently do not induce any. In the former, the Curie and second transition temperatures vary systematically, but differently, with $x$. Our results imply that there is a definite proportionality between the $x$ dependences of the second transition temperature and the $f-d$ hybridization strength estimated qualitatively from the absorption spectra.

\end{abstract}

\maketitle

\section{INTRODUCTION}

The C15 Laves phase compound CeFe$_2$ is known to be an unstable ferromagnet with the existence of antiferromagnetic (AFM) spin fluctuations in ferromagnetic (FM) ground state. \cite{Kennedy90, Paolasini03} This instability has been extensively studied with the help of impurity substitutions at the Fe site. A very small amount ($\sim$ 5 $\%$) of certain impurities, specifically Co, Ru, Ir, Al, Os, Re, Ga, Si, is known to cause a total loss of ferromagnetism at temperatures lower than the CeFe$_2$ Curie temperature (T$_{\rm C}~\sim$ 230 K). \cite{royprb, Roy89, Chaboy00, Roy04, Haldar08, Vershinin14} This second, FM-AFM, transition has been reported to be of first-order nature. \cite{Chatt03, Manekar01, Sokhey04} For certain other impurities, viz., Mn, Ni, Cu, Rh, Pd, and Pt, however, no such additional magnetic phase transition occurs. \cite{Chaboy00, Kennedy90} Various experimental tools have been employed to know the true magnetic state of Ce sub-lattice and the instabilities related to the Fe sub-lattice. Theories in the literature suggest that the instability of the FM state arises through a competition between the ferromagnetic Fe 3\emph{d}-Fe 3\emph{d} interaction, and the antiferromagnetic interaction due to the Ce 4$f$ - Ce 5$d$ - Fe 3$d$ hybridization. \cite{Wang12}

It is quite intriguing to find a possible connection between the occurrence or non-occurrence, and the composition-dependent behaviour of the second transition (if present), on the one hand, and the strengths of the $d-d$ interaction and $f-d$ hybridization on the other. There exists one experimental report by Chaboy {\it {et al.}}, \cite{Chaboy00} wherein they study the $f-d$ hybridizations in Ce(Fe$_{1-x}$Co$_x$)$_2$ using Ce L$_{1,3}$-, Fe K- and Co K-edge x-ray absorption spectroscopy (XAS) and x-ray magnetic circular dichroism. They, however, address a different issue. They correlate an anomalous magnetic behavior of Ce(Fe$_{1-x}$Co$_x$)$_2$, best understandable as the non-linear (somewhat oscillatory) $x$-dependence of the T$_{\rm C}$, with the Ce valence and the $f-d$ hybridization, and found that neither of the two is responsible for the anomaly. The behaviour of the second transition was not looked at in the report. In a step forward in this direction, we recently reported an extensive systematic computational investigation of the behaviour of the $f-d$ hybridization in representative Ce(Fe$ _{0.75} $M$ _{0.25} $)$ _2$ compounds across 3\textit{d}, 4\textit{d}, 5\textit{d} and post-transition impurity series. \cite{Das16} The relevant outcome of this computational study has been that the $f-d$ hybridization is the strongest for the Mn group impurity in the period, and gets weakened on either side of it. 

Another consideration which possibly could help one seek the origin of the occurrence or non-occurrence of the impurity-induced second transition could be the crystal structure of the end-compound CeM$_2$, along with the solid solubility of M in CeFe$_2$. A survey of the literature cited above reveals that out of the existing second transition inducing impurities, Al, Co, Ru, Os and Ir possess the same cubic Laves phase structure (space group 277) as CeFe$_2$, while CeSi$_2$ (space group 141) and CeGa$_2$ (space group 191) are of different lattice structures, and CeRe$_2$ does not even exist thermodynamically. Among the impurities not causing the second transition also, some of the end-compounds (CeNi$_2$, CeRh$_2$ and CePt$_2$) are isostructural with CeFe$_2$ and some different (CeCu$_2$ - space group 74; CeSb$_2$ - space group 64). So, the crystal structure of the end-compound CeM$_2$ has no correlation with the second transition. Nevertheless, we have chosen three impurities Cr, Ag and Au such that either there exists no CeM$_2$ end-compound (CeCr$_2$) thermodynamically or, if it exists, it is not a Laves phase (CeAg$_2$, CeAu$_2$: space group 74) compound. From the computation point of view, \cite{Das16} the $f-d$ hybridization in these pseudobinaries are supposed to be weaker than that in CeFe$_2$. These impurities have hitherto not been studied.

The present work, thus, comprises of a two-fold objective: (i) to explore the occurrence and behavior of the second transition in pseudobinaries Ce(Fe$_{1-x}$M$_x$)$_2$ with unexplored impurities Cr, Ag and Au, and (ii) to seek any correlation between the behavior of the second transition and the $f-d$ hybridization. In the following we will show that for Ag and Au impurities, although the hybridization as calculated using XAS decreases in agreement with the theoretically predicted behaviour, there is no second phase transition. In the case of Ce(Fe$_{1-x}$Cr$_x$)$_2$, however, the second (FM-AFM) transition exists, and the $f-d$ hybridization nicely follows the concentration dependence of the FM-AFM transition temperature. The work, this way, is another step forward to find origin of the occurrence or non-occurrence of the second transition in Fe substituted CeFe$_2$.

\section{EXPERIMENT}

Polycrystalline CeFe$_2$, and Ce(Fe$_{1-x}$Cr$_x$)$_2$ ($x$ = 0.01, 0.02, 0.025, 0.03, 0.04), Ce(Fe$_{0.96}$Ag$_{0.04}$)$_2$, and Ce(Fe$_{0.97}$Au$_{0.03}$)$_2$ alloys were prepared by arc melting high-purity ($\geq 99.9 ~\%$) elemental metals in an argon atmosphere. The alloy buttons were remelted thrice to ensure homogeneity. The buttons were then sealed separately in evacuated quartz tubes and annealed in a controlled manner as suggested by Roy \textit{et al.}. \cite{Roy89} Energy dispersive analysis of x-rays (EDAX) was performed on each sample to have an estimation of the atomic percentages of the elements present using a JEOL JEM-2100 scanning electron microscope. Structural analyses of the samples were performed at room temperature by powder x-ray diffraction (XRD) using a Cu-K$_\alpha$ source from a Philips X'Pert MRD x-ray diffractometer. The samples were further analysed by x-ray photoelectron spectroscopy (XPS) at room temperature using a PHI 5000 Versa Probe II system. The XPS data were recorded under ultra-high vacuum condition with a base pressure of  $\sim ~5 \times 10^{-10}$ mbar using an Al-K$_\alpha$ source, a hemispherical analyser and a multichannel detector. In order to reduce the surface effects, clean sample surfaces were first obtained by cleaving the polycrystals before inserting into the UHV chamber, and then sputtering the surface using Ar$^+$ ions before the data collection.

The dc magnetization measurements were performed using either a physical property measurement system from Cryogenic or a superconducting quantum interference device from Quantum Design. X-ray Absorption Spectroscopy (XAS) experiments were performed at the Ce M$_{4,5}$ edges using the soft x-ray beam in total electron yield mode from the bending magnet port BL-1 of the INDUS-2 ring at Raja Ramanna Centre for Advanced Technology, Indore, India. Samples were cleaved before transferring them into the UHV XAS chamber. 

\section{RESULTS AND DISCUSSION}

\subsection{X-ray diffraction}

Figure 1(a) shows the XRD patterns of all the samples. The patterns are compared with  Joint Committee on Powder Diffraction Standards (JCPDS) file (1999) and indicate that the samples are in a single-phase MgCu$_2$-type cubic Laves phase structure, with only a very weak cerium oxide peak at 2$\theta$ $\sim$ 28$^{\circ}$ and 47$^{\circ}$ in some compounds. The appearance of the minor cerium oxide is inevitable due to the chemically very reactive surface characteristics \cite{Fukuda01} of the pseudobinaries. However, a peak area based quantification suggests that the amount of the oxide does not exceed 5 $\%$ and hence would not influence the alloy properties substantially.

\begin{figure}[h]

\centering
\begin{subfigure}{1\textwidth}
  \centering
    \includegraphics[width=4 in]{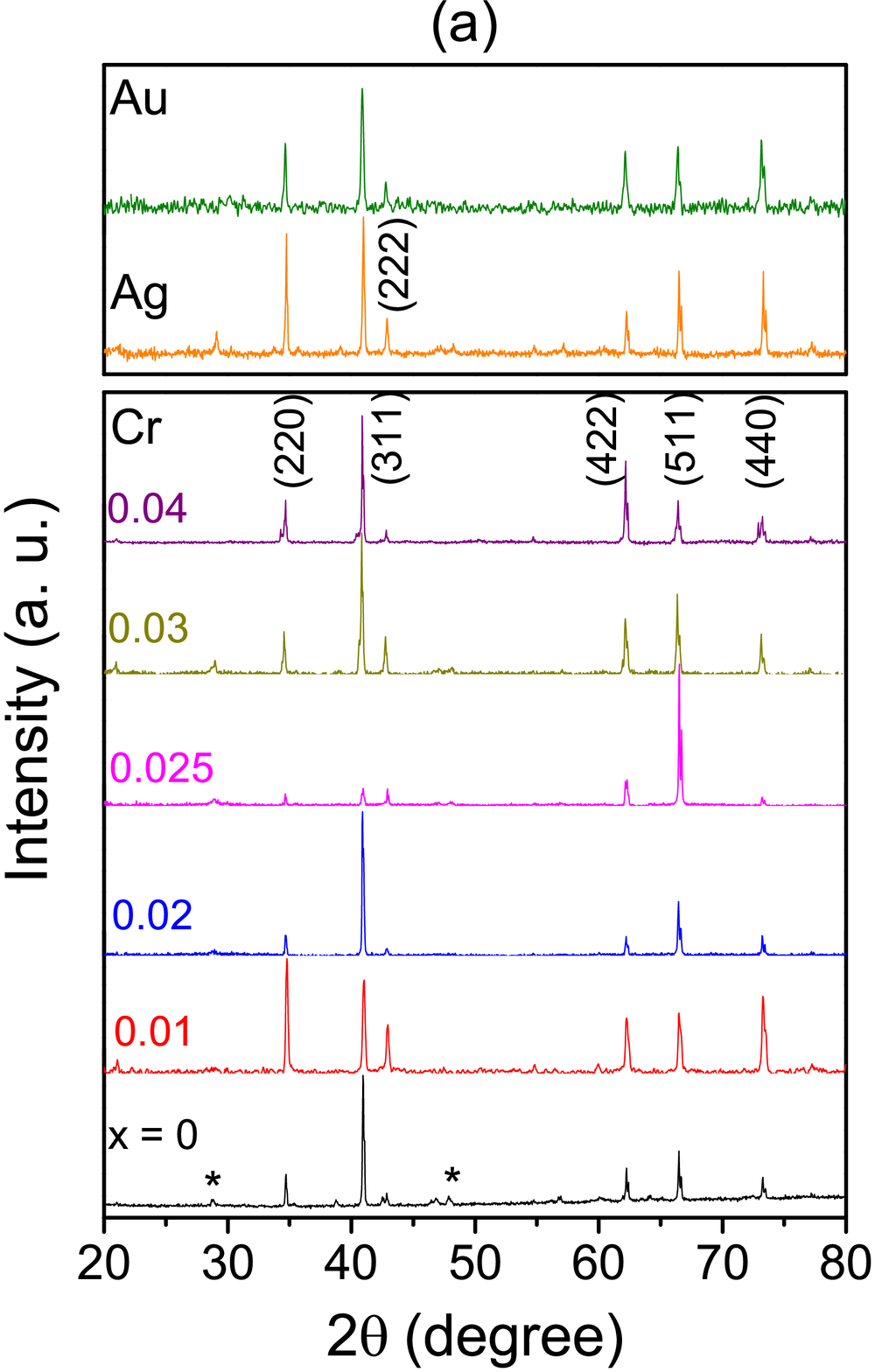}
\end{subfigure}%

\begin{subfigure}{1\textwidth}
  \centering
    \includegraphics[width=4 in]{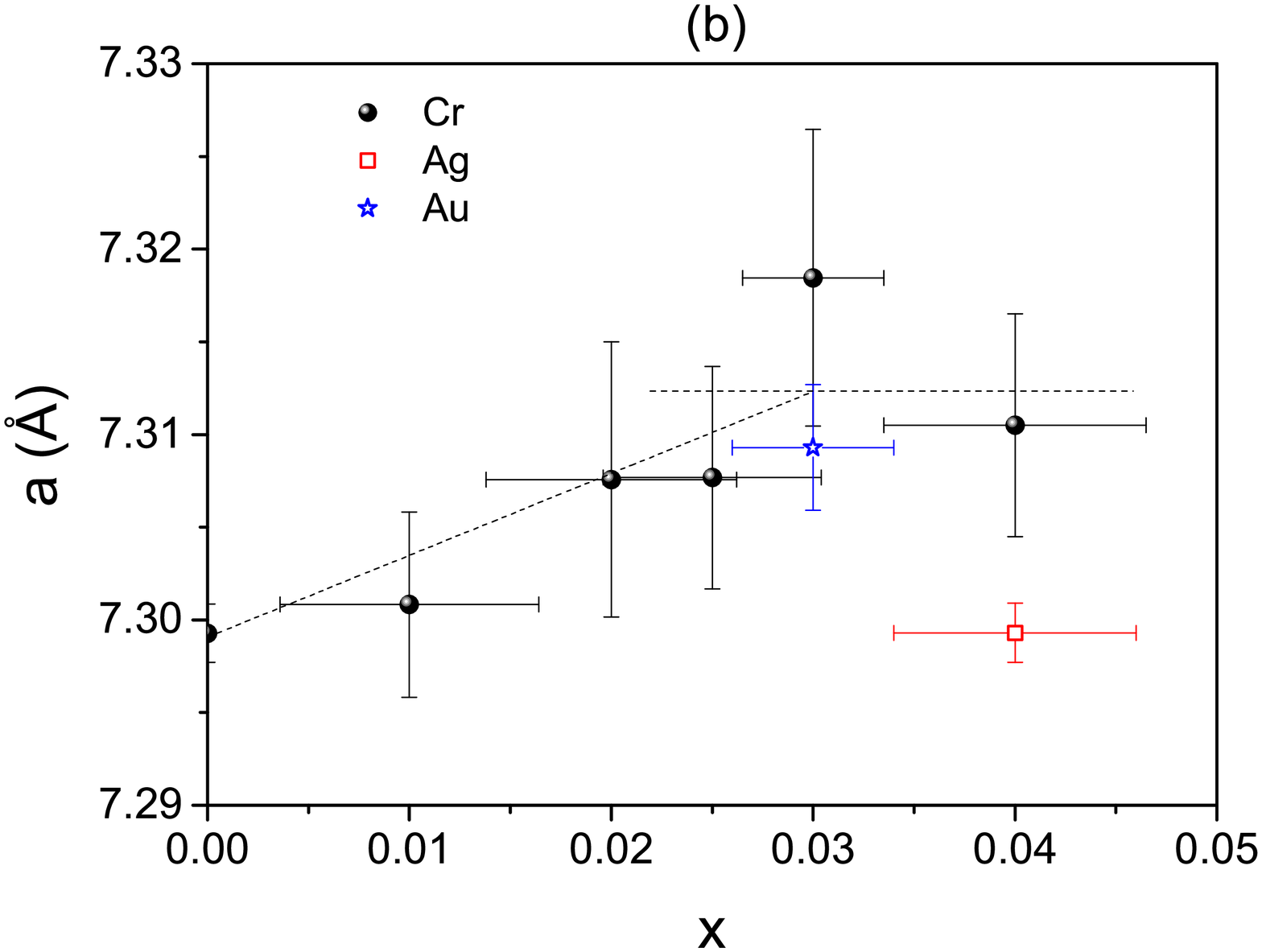}
\end{subfigure}%

\caption{(a) XRD pattern of Ce(Fe$_{1-x}$M$_x$)$_2$ samples. Top panel - M = Ag and Au; bottom panel - M = Cr for different $x$. (b) Variation of lattice constant with $x$ for different M.}

\label{figureone}

\end{figure}

Figure 1(b) displays the variation of lattice constant $a$ of the alloys with composition $x$. The $x$ errors have been determined by considering the EDAX estimated compositions and quantitative compositional analyses of high-resolution Fe and M XPS spectra (not shown). The average lattice constants and corresponding errors have been determined by fitting each prominent XRD peak of an alloy with the Bragg's law, collecting the different $a$ values and then averaging them out. The average lattice constant of CeFe$_2$ is thus found to be 7.299 $\pm$ 0.002 \AA, and agrees well with the existing literature value. \cite{Forsthuber90} Even with the large error bars, the following can very tentatively be inferred from the figure.  For Cr impurities, the lattice constant increases linearly up to $x~\sim$ 0.03 and then becomes constant. The increase must be due to the larger atomic radius of Cr (1.39 \AA) than that of Fe (1.32 \AA), which is being substituted. However, a comparison of the lattice constant increase with Vegard's law is not possible because the end-compound CeCr$_2$ does not exist and hence its lattice constant is not known. The apparent attainment of constant $a$ value for $x~>$ 0.03 indicates that the solubility limit of Cr in CeFe$_2$ may be $x~\sim$ 0.03. In case of Au also, the increase in $a$ is in agreement with the larger atomic radius of Au (1.36 \AA) than of Fe. In case of Ag, however, there is no change in the lattice constant at 4 atomic $\%$ Ag. Possibly, Ag has no solid solubility in CeFe$_2$.

\subsection{X-ray photoelectron spectroscopy}

Figure 2(a) displays the Ce 3\emph{d} core-level spectrum along with its deconvolution. As can be seen, each spin-orbit split line-shape consists of three peaks, viz., 3\emph{d}$^{9}$4\emph{f}$^{0}$, 3\emph{d}$^{9}$4\emph{f}$^{1}$ and 3\emph{d}$^{9}$4\emph{f}$^{2}$, a typical feature of the strongly hybridized CeFe$_2$, as also reported in the literature. \cite{Konishi00} We refer to them as $f_P^0$, $f_P^1$ and $f_P^2$ peaks, respectively, P denoting photoemission. There are also components corresponding to CeO$_2$, and an area based quantitative analysis puts the CeO$_2$ concentration at $\sim 8 \%$, not much different from the XRD results. However, CeO$_2$ being non-magnetic, it would not affect the present results and findings to any extent.

\begin{figure}[h]

\centering
\begin{subfigure}{.5\textwidth}
  \centering
    \includegraphics[width=4.5 in]{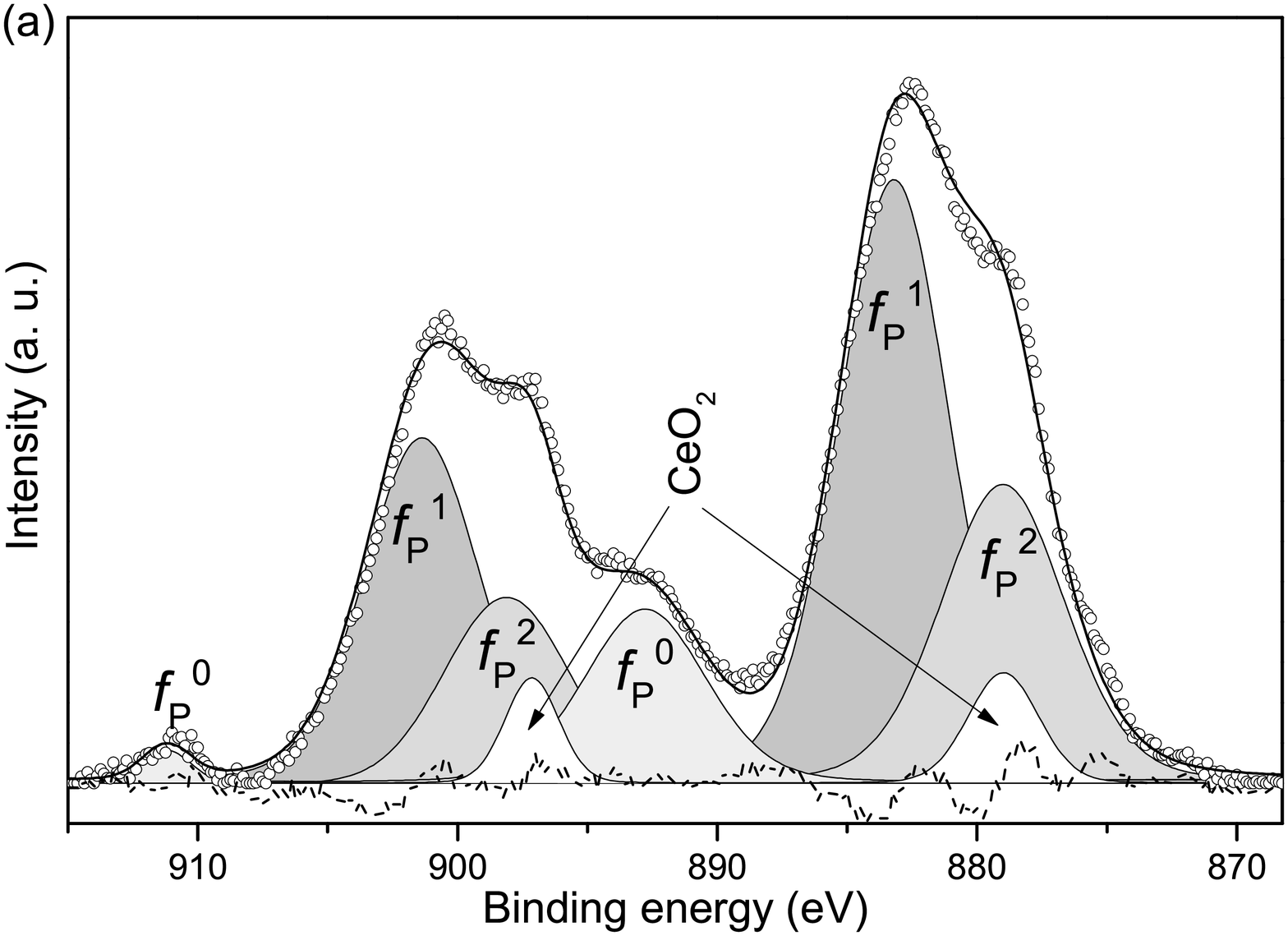}
\end{subfigure}%

\begin{subfigure}{.5\textwidth}
  \centering
    \includegraphics[width=4.5 in]{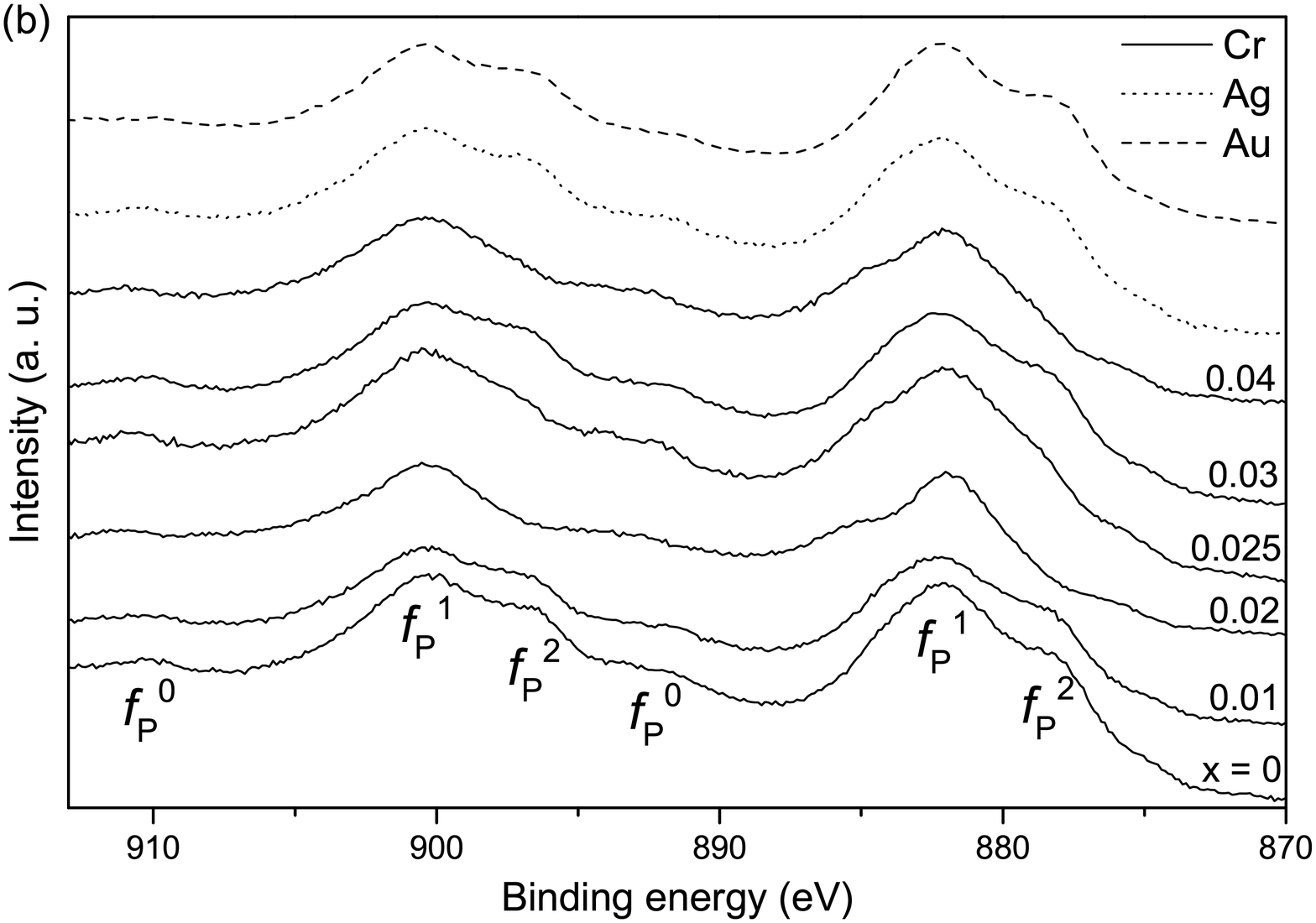}
\end{subfigure}%

\caption{(a) Ce 3d XPS spectra of CeFe$_2$ with the deconvoluted peaks. Open circles represent the data. the thicker line is is the fit, and the thinner lines separate the deconvoluted peaks. The dotted curve is the difference between the data and the fit. (b) Ce 3d XPS spectra of Ce(Fe$_{1-x}$Cr$_x$)$_2$ for all $x$ and M.}

\label{figuretwo}

\end{figure}

The Ce 3\emph{d} core-level spectra of all the samples are shown in Fig. 2(b). Deconvolutions (not shown) of the spectra suggest that all these contain the same features, although with varying amounts of CeO$_2$. According to the deconvolutions, all the $f_P^0$, $f_P^1$ and $f_P^2$ peaks occur at the same respective positions for all the samples, suggesting that the oxidation states of Ce remain unaltered on the introduction of impurities. Although there exists a prescription of estimating the strength of $f-d$ hybridization, or at least the change in this strength on impurity substitution, via the ratio $f_P^0/(f_P^0+f_P^1)$ of the area of the $f_P^0$ peak to the sum of the areas of the $f_P^0$ and $f_P^1$ peaks, \cite{Chaboy00, Fuggle80, Fuggle83} the impurity concentrations are so small that the changes would be negligible. The estimation with the area ratio becomes even more erroneous because of the spectrum to spectrum variation in the determination of the background, which has to be subtracted before finding the areas. For this reason, we do not interpret the XPS spectra further and limit its usage just to demonstrate that the samples are satisfactorily good for further analyses.

\subsection{Magnetization}

Figure 3 shows the temperature dependence of zero-field-cooled (ZFC) magnetizations (M-T) of CeFe$_2$, Ce(Fe$_{0.96}$Au$_{0.03}$)$_2$ and Ce(Fe$_{0.97}$Ag$_{0.04}$)$_2$ samples in the temperature range 5 K to 300 K and in an external field 500 Oe for the first two samples and 50 Oe for the last one. The parmagnetic (PM) to ferromagnetic phase transition can be seen to occur at T$_{C1} \sim$ 225 K for all the samples. Further, the FM state of the Ag and Au substituted samples continues to exist down to 5 K, as in CeFe$_2$. Since the concentrations of all the reported second transition inducing impurities lie in this range, we believe that Ag and Au would not cause the FM-AFM transition at any impurity concentration.

\begin{figure}[h]

  \centering
    \includegraphics[width = 5 in]{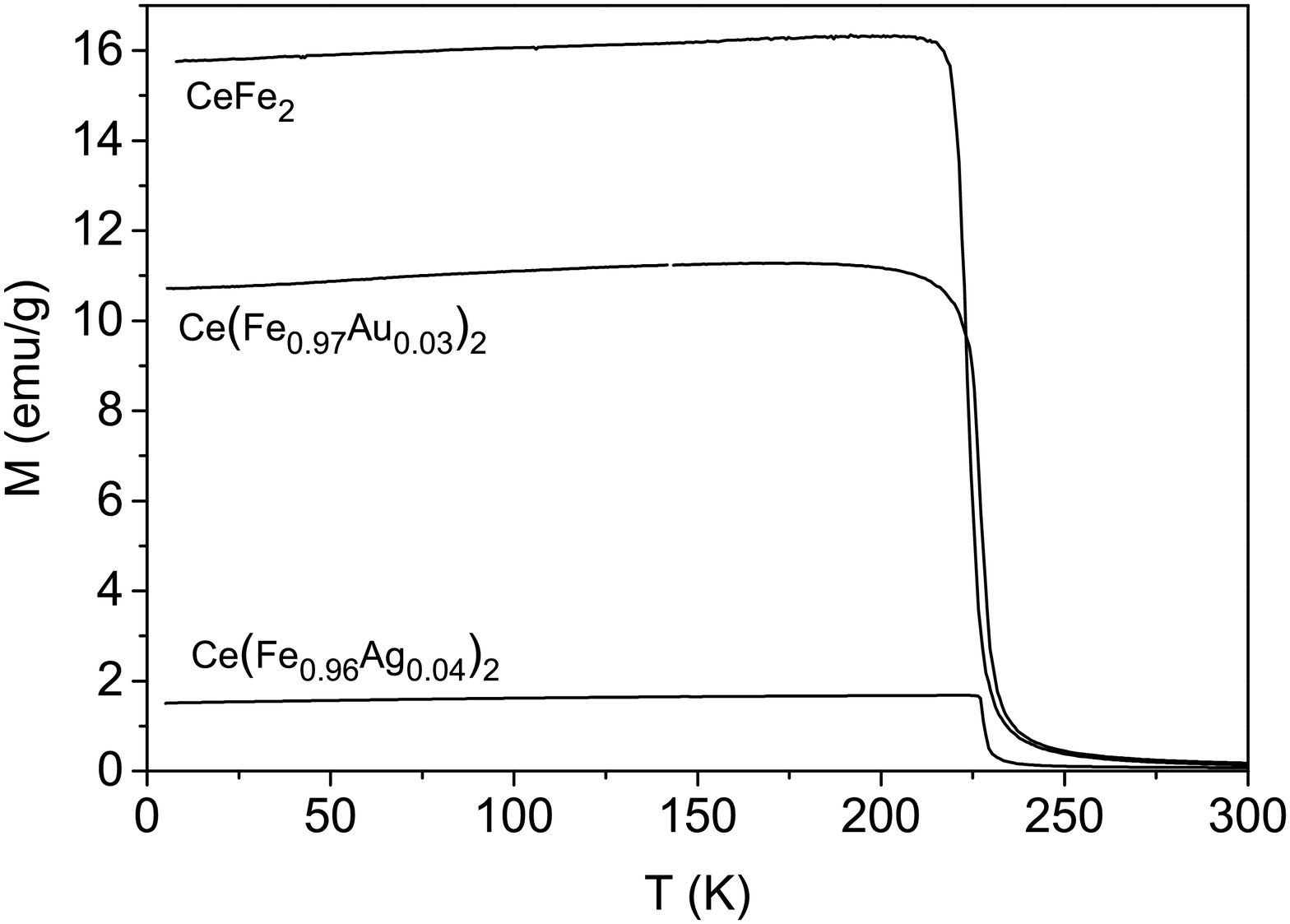}

\caption{Temperature dependence of magnetization for CeFe$_2$, Ce(Fe$_{0.97}$Au$_{0.03}$)$_2$ and Ce(Fe$_{0.97}$Ag$_{0.04}$)$_2$ compounds under an external field of 500 Oe for the former two and 50 Oe for the latter.}

\label{figurethree}

\end{figure}  

The ZFC magnetizations of Ce(Fe$_{1-x}$Cr$_x$)$_2$ samples, including that of CeFe$_2$, in the same temperature range and fields as above, are shown in Fig. 4(a). From the figure, the T$_{C1}$ can be seen to vary with $x$. Such $x$-dependence of T$_{C1}$ has been reported earlier for Ga, Co and Si impurities in CeFe$_2$. \cite{Haldar08, Vershinin14, Zhang94} Apart from this, a second transition can be seen at lower temperatures for $x \geq {0.01}$, with an $x$-dependent transition temperature T$_{C2}$ below 50 K. However, the magnetization below T$_{C2}$ does not go to zero, suggesting that the FM phase coexists with the new phase at lower temperatures at 500 Oe field. Further, the second transition for $x$ = 0.01 sample is quite feeble. Coexistence of FM and AFM phases, leading to a canted spin structure, has been reported earlier in Ce(Fe$_{0.98}$Ru$_{0.02}$)$_2$ \cite{Kennedy90} and Ce(Fe$_{0.95}$Co$_{0.05}$)$_2$. \cite{Fukuda01} The second transition has been reported to be of the first order for these systems. \cite{Kennedy90, Fukuda01} An indication of the low temperature phase in Ce(Fe$_{1-x}$Cr$_x$)$_2$ ($x \geq {0.01}$) being AFM can be found in Fig. 4(b), wherein the M-T plots under both ZFC and field-cooled-warming (FCW) conditions are displayed for various external fields. The figure suggests that the second transition gets gradually suppressed with the increase of magnetic field, and the field as high as 40 kOe is able to suppress the low temperature transition completely. Such observations have also been reported for Ga-doped CeFe$_2$ compounds \cite{Haldar08} and have been linked with the suppression of the second transition by the field.

\begin{figure}[h]

\centering
\begin{subfigure}{.5\textwidth}
  \centering
    \includegraphics[width=5 in]{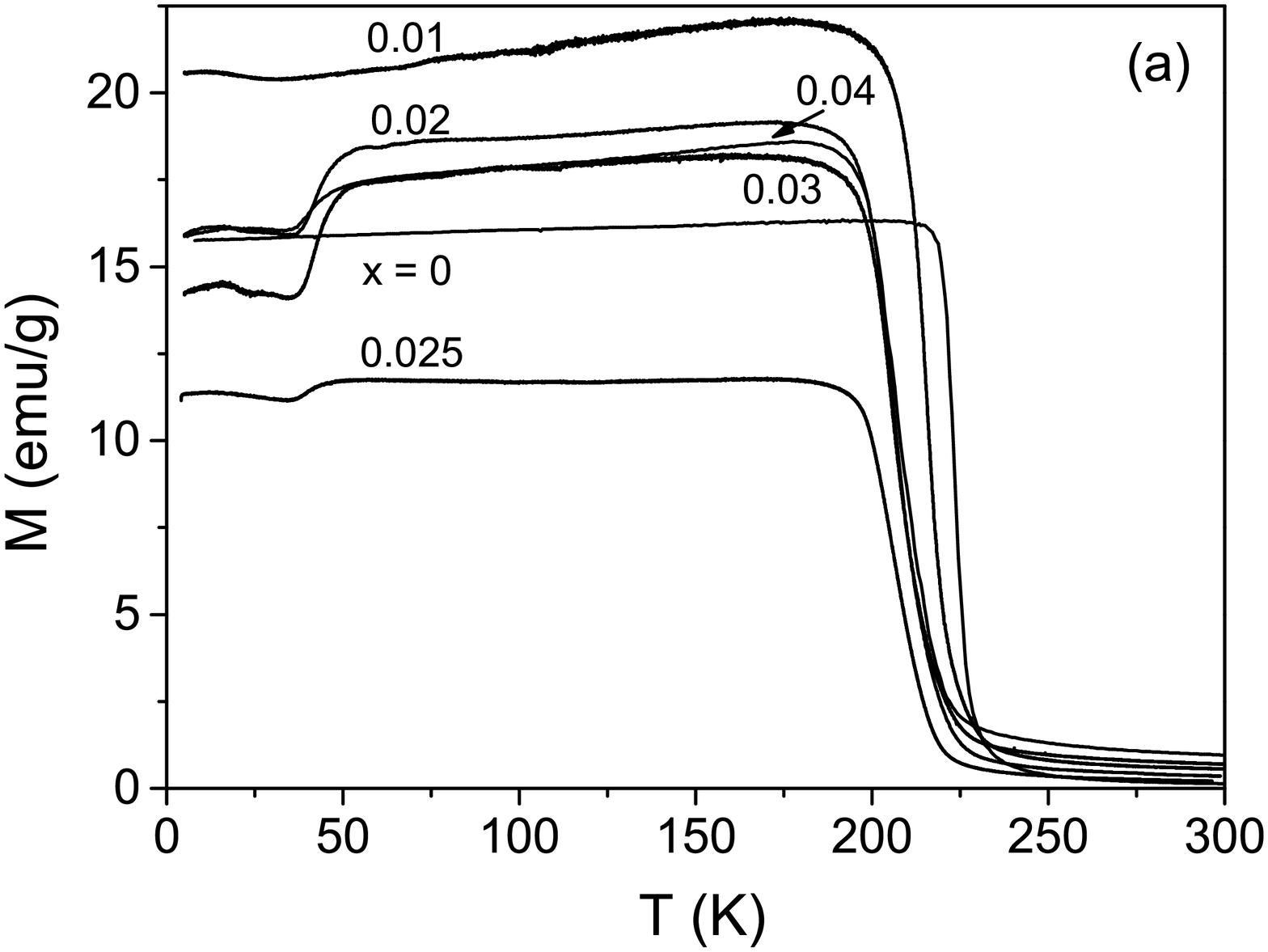}
\end{subfigure}%

\begin{subfigure}{.5\textwidth}
  \centering
    \includegraphics[width=5 in]{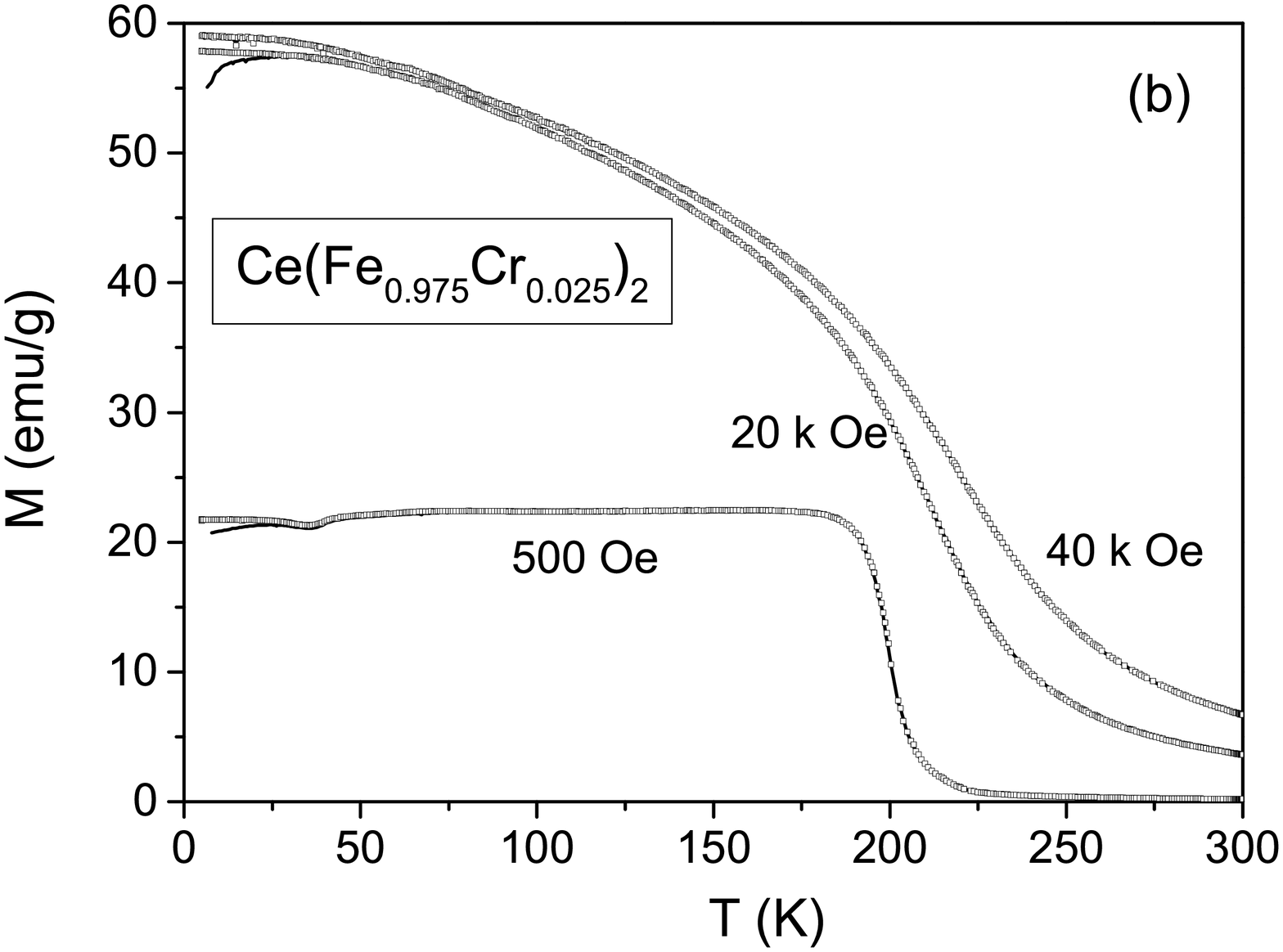}
\end{subfigure}%

\caption{(a) M-T curves for CeFe$_2$ under 500 Oe external field. (b) M-T curves for Ce(Fe$_{0.975}$Cr$_{0.025}$)$_2$ under different external fields. Symbopls and lines represent FC and ZFC data, respectively.}

\label{figurefour}

\end{figure} 

The magnetic nature of the low temperature phase in Ce(Fe$_{1-x}$Cr$_x$)$_2$ can further be inferred from the hysteresis (M-H) curves measured at 2 K and shown in Figs. 5(a) - 5(d). Figure 5(a) suggests that CeFe$_2$ is a normal ferromagnet showing saturation already at very small fields. This behaviour of CeFe$_2$ and its saturation magnetization 55.8 emu/g, equalling a magnetic moment of 2.48 $\mu_B$/ f.u., is in agreement with the literature. \cite{Eriksson88} The Ce(Fe$_{0.99}$Cr$_{0.01}$)$_2$ sample also shows more or less the same behaviour since the second transition as observed in the corresponding M-T curve is weak. For $x \geq {0.02}$ samples, which show a clear second transition in the corresponding M-T curve, five-quadrant M-H isotherms have been taken (Figs. 5(b) - 5(d)) at 2 K with the minimum field sweep rate 200 Oe/min. In each of these, the magnetization is linear with the field at very low fields, which is indicative of the presence of an AFM phase at this temperature, as reported also by others. \cite{Kennedy90, Haldar08} The domains start lying parallel to the field during the increment of the field further, and the FM loop opens up at higher fields. Such field induced transitions have been reported for Ru-, Re- and Ga-substituted CeFe$_2$. The loop openings, thus strongly suggest that the low temperature phase, coexistent with the FM phase, is AFM. The asymmetry in the data between the  increasing and decreasing field cycles could be due to the supercooling effect associated with a first-order phase transition. \cite{Roy04, Chatt03} The relatively sharp multiple magnetization step observable for $x$ = 0.025 (Fig. 5(c)) has a marked similarity with Re-, Ru-, Ga-substituted CeFe$_2$ compounds and with some Perovskite and GMR manganites. \cite{Roy05, Haldar08, Mahen02, Levin02} The step-type behaviour can be linked with the presence of a martensitic strain in the system, \cite{Roy05, Haldar08, Mahen02, Levin02} but this aspect is not important for the present study. The occurrence of such steps has also been suggested to be due to the thermal and magnetic history of the samples. \cite{Manekar01, Singh02} Just in order to check this, two-loop M-H isotherms were also recorded for the Ce(Fe$_{0.975}$Cr$_{0.025}$)$_2$ sample without field cycling and in the sequence 0-5-0-5-0 T at the sweep rate of 200 Oe/min. The data are presented in Fig. 6. The isotherms in this case are found to be different from the previous run in that the magnetization attains near saturation value at relatively lower fields, and demonstarte the influence of magnetic history in determining the step or smooth behaviour of the transition.

\begin{figure}[h]

\centering
\begin{subfigure}{.5\textwidth}
  \centering
    \includegraphics[width = 3 in]{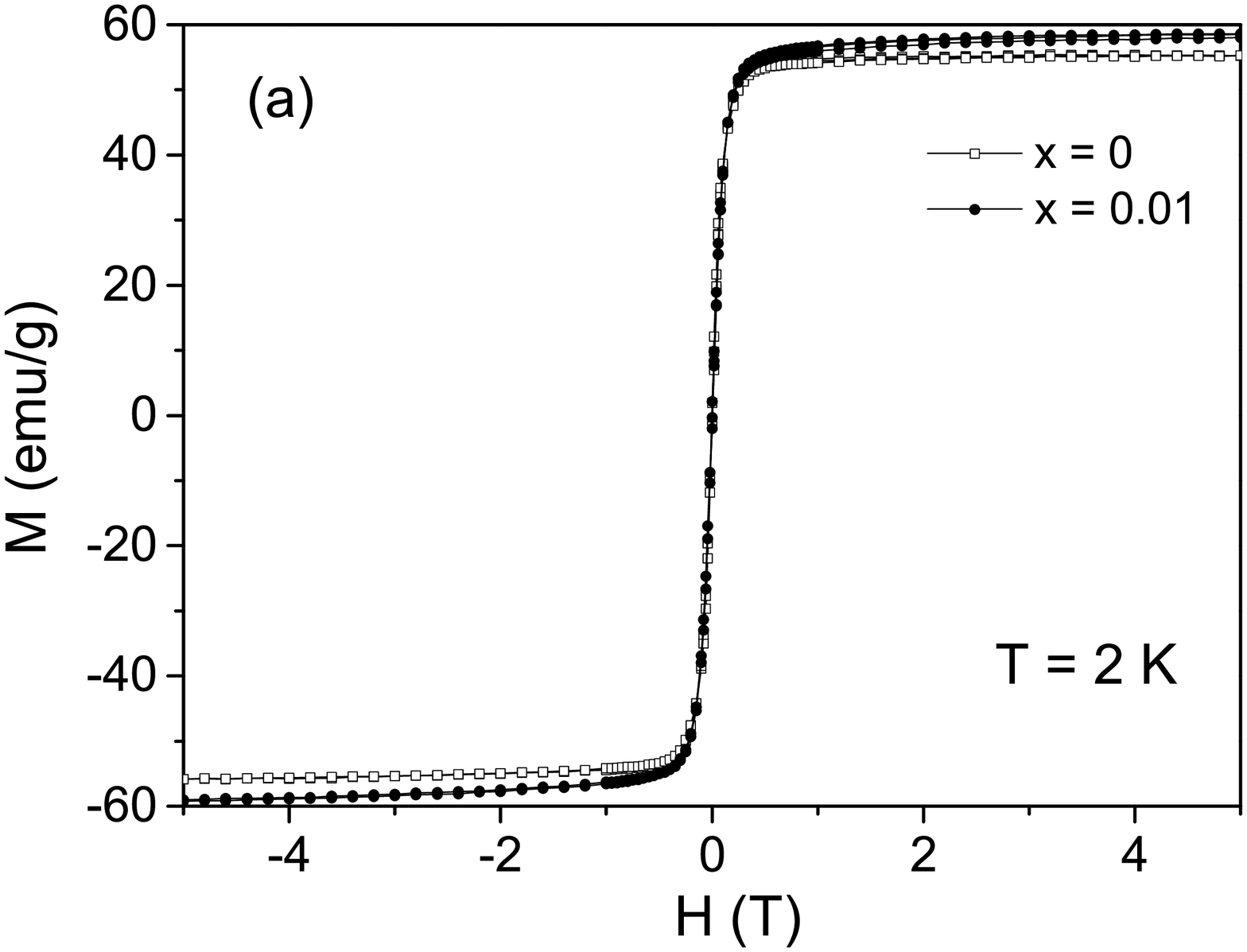}
\end{subfigure}%

\begin{subfigure}{.5\textwidth}
  \centering
    \includegraphics[width = 3 in]{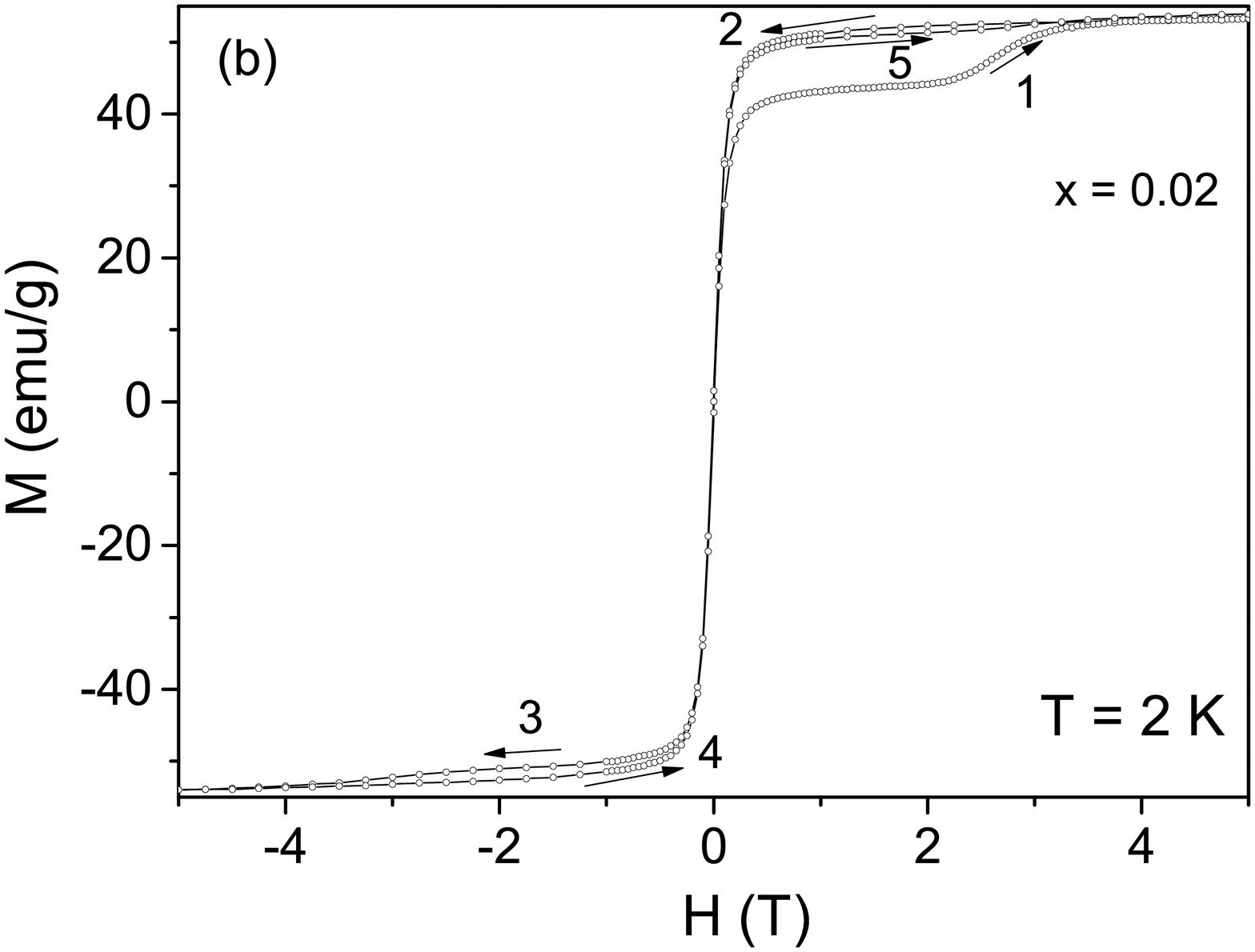}
\end{subfigure}%

\begin{subfigure}{.5\textwidth}
  \centering
    \includegraphics[width = 3 in]{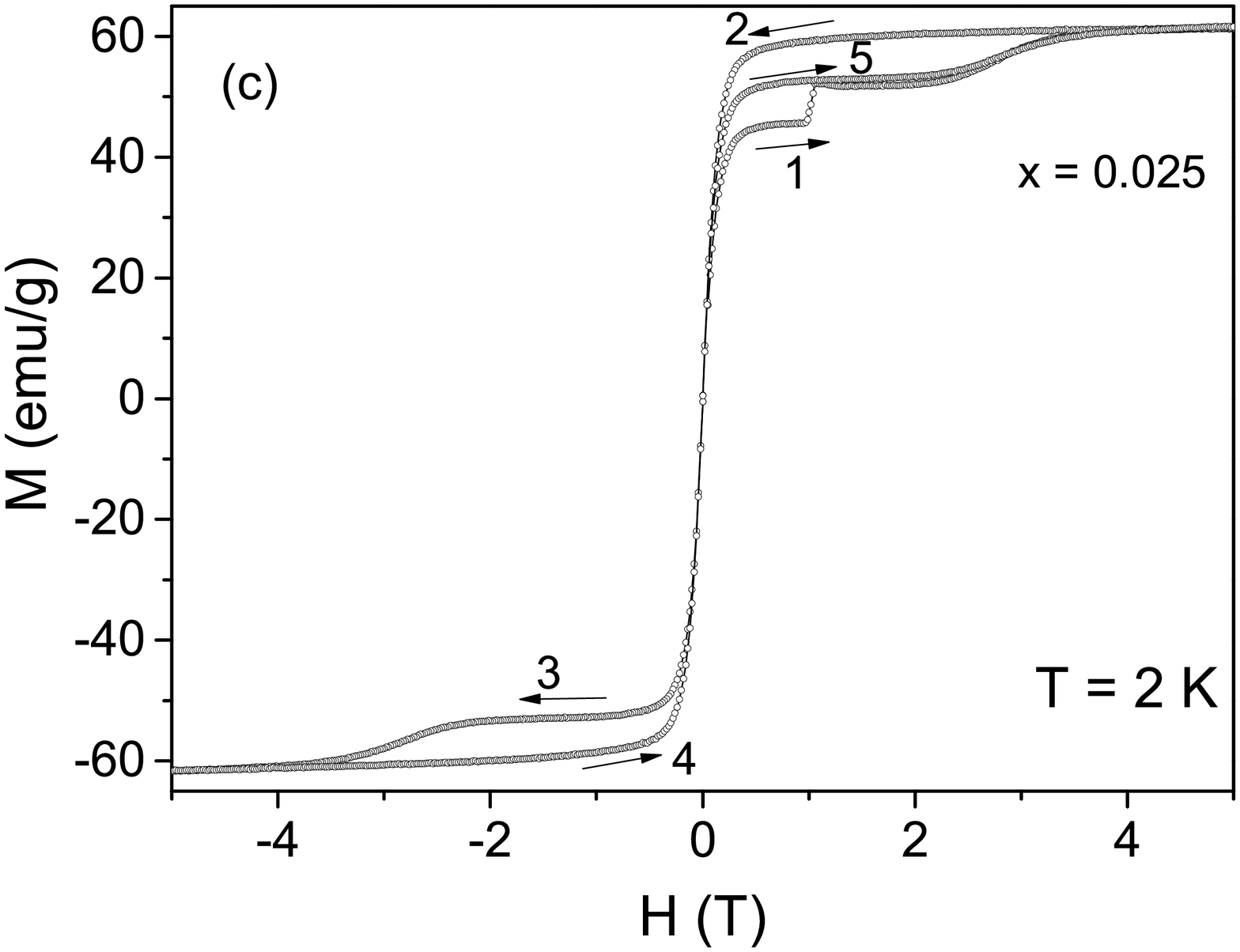}
\end{subfigure}%

\begin{subfigure}{.5\textwidth}
  \centering
    \includegraphics[width = 3 in]{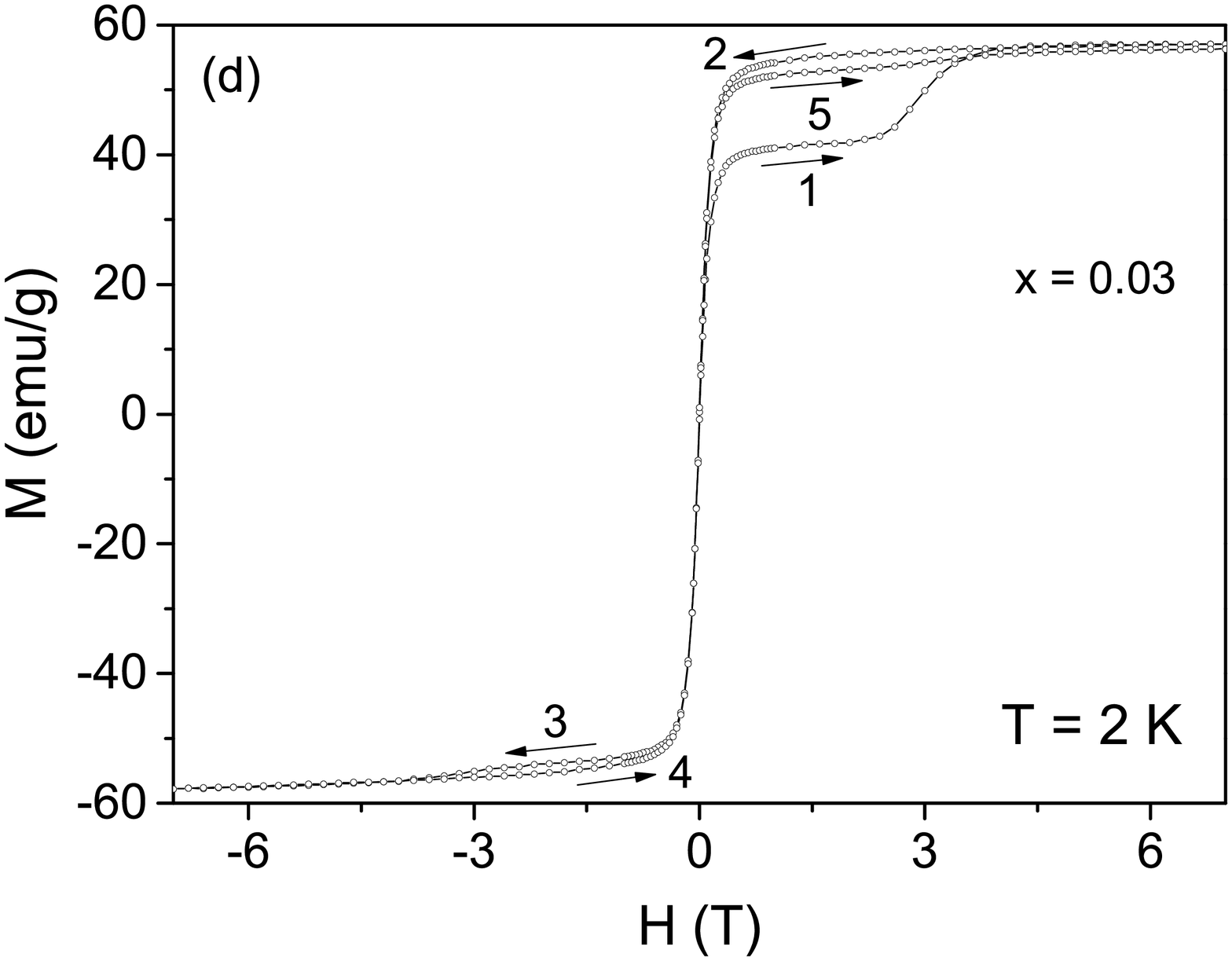}
\end{subfigure}%

\caption{M-H hysteresis curves for CeFe$_2$ and Ce(Fe$_{0.99}$Cr$_{0.01}$)$_2$ (a). Five-quadrant M-H isotherms of Cr series for $x$ = 0.02 (b), 0.025 (c) and 0.03 (d).}

\label{figurefive}

\end{figure}

\begin{figure}[h]

\centering
  \centering
    \includegraphics[width = 4 in]{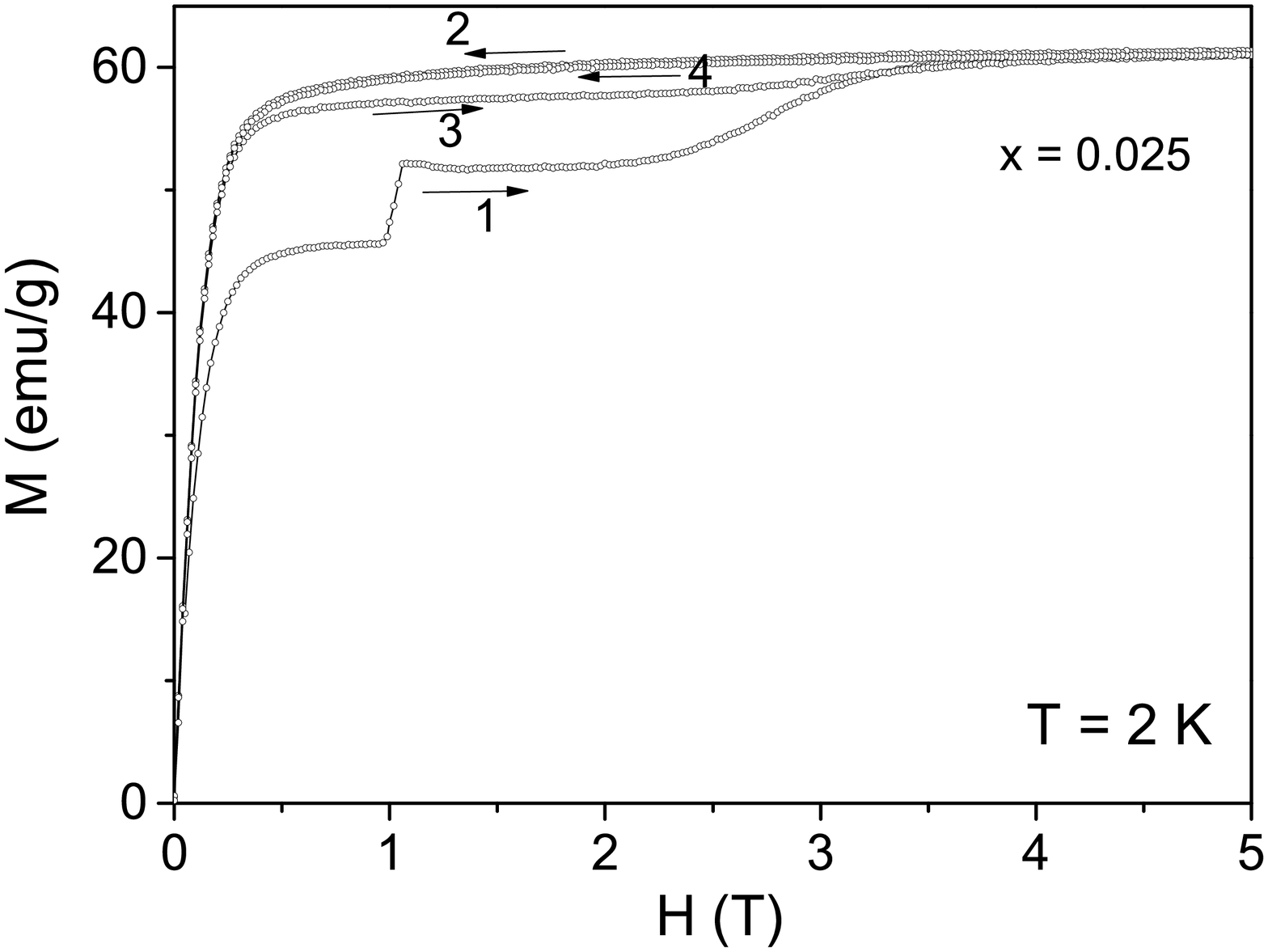}

\caption{M-H isotherms for Ce(Fe$_{0.975}$Cr$_{0.025}$)$_2$ compound without field cycling.}

\label{figuresix}

\end{figure}

We chose the Ce(Fe$_{0.975}$Cr$_{0.025}$)$_2$ sample for finding out the natures of the two transitions and recorded Arrott plots (H/M versus M$^2$) near T$_{C1}$ and T$_{C2}$. The plots are shown in Figs. 7(a) and 7(b), respectively. Positive slopes possessed by all the curves near T$_{C1}$ (Fig. 7(a)) suggest that the first (PM to FM) transition is of second order, \cite{Bustingorry16} in consistence with the literaure. The change of curvature happens at around 210 K, putting the T$_{C1}$ value for this pseudobinary at $\sim$ 210 K. However, the Arrott plots near T$_{C2}$ have a tendency to have a negative slope beyond 40 K. This clearly indicates that the second (FM to AFM) transition is of first order nature \cite{Bustingorry16} and that the T$_{C2}$ value for this pseudobinary is $\sim$ 40 K. 

\begin{figure}[h]

\centering
\begin{subfigure}{.5\textwidth}
  \centering
    \includegraphics[width=5in]{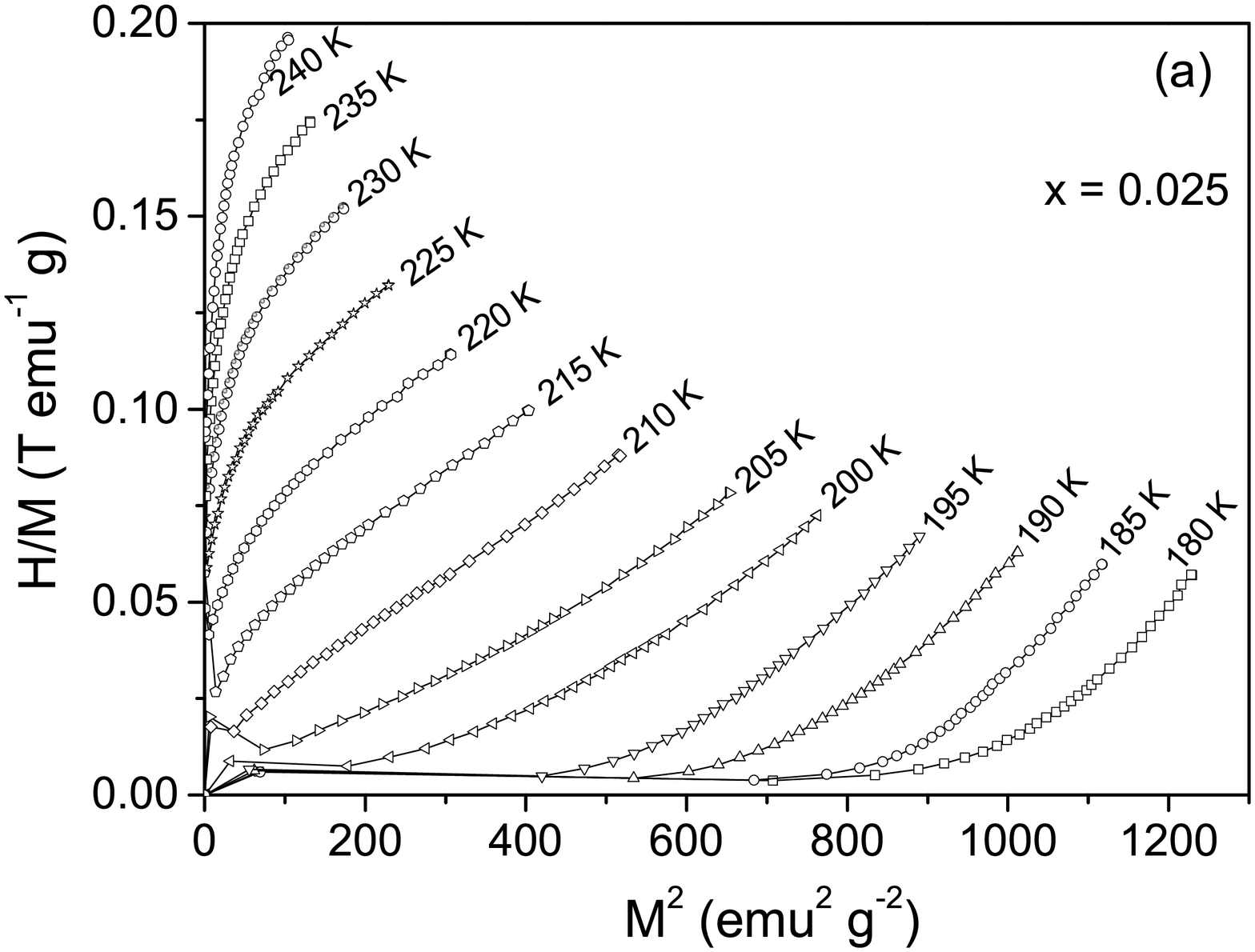}
\end{subfigure}%

\begin{subfigure}{.5\textwidth}
  \centering
    \includegraphics[width=5in]{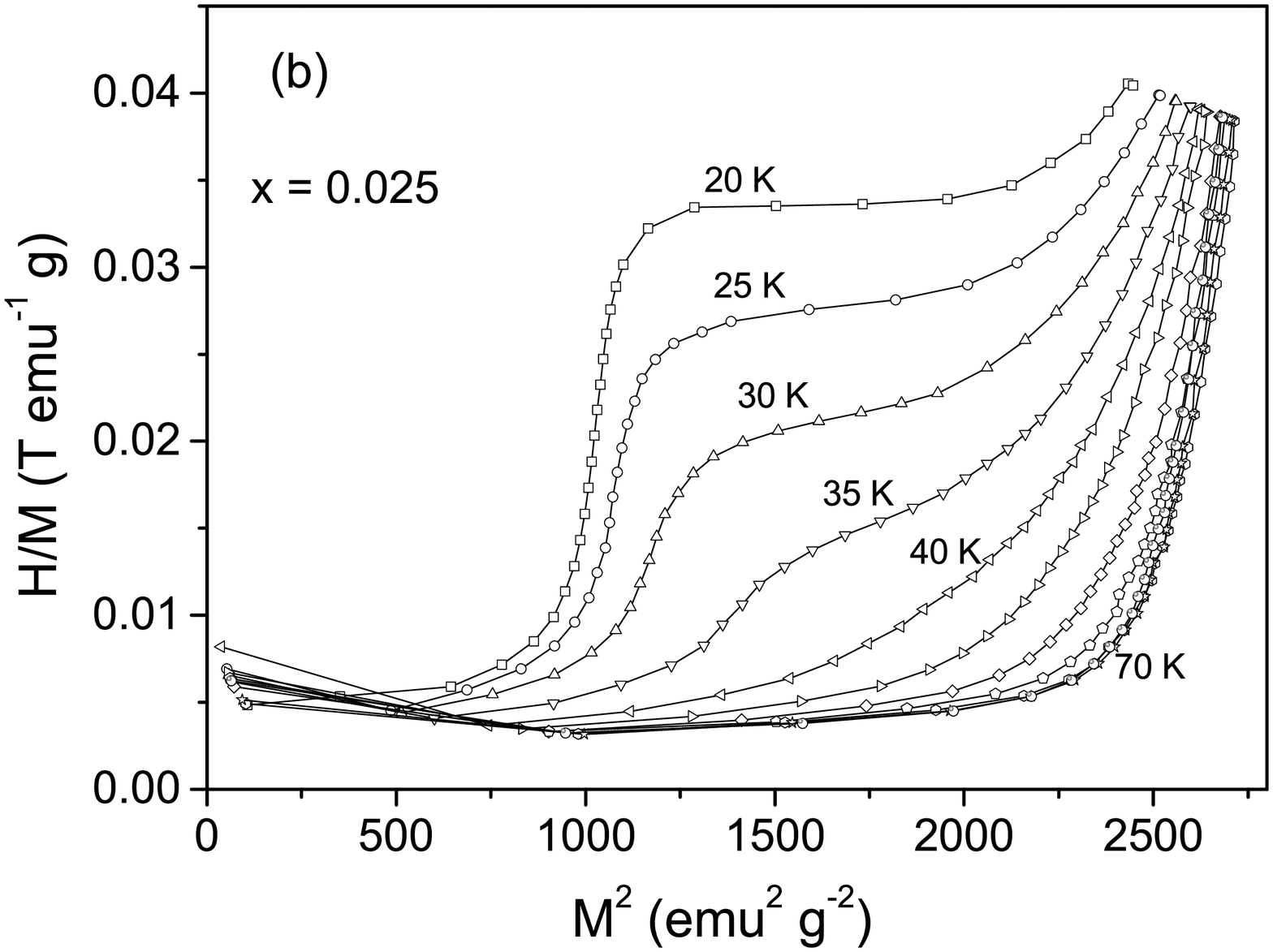}
\end{subfigure}%

\caption{Arrott plots for Ce(Fe$_{0.975}$Cr$_{0.025}$)$_2$ compound near first (a) and second (b) transitions.}

\label{figureseven}

\end{figure}

Based on the above discussion, a T - $x$ magnetic phase diagram has been drawn in Fig. 8. For Cr substitutions, the phase diagram can be divided into three regions separated by T$_{C1}$ and T$_{C2}$ curves. As can be seen from the figure, T$_{C1}$ falls visually exponentially from $\sim$ 225 K to $\sim$ 210 K on increasing the Cr concentration and separates the PM and FM regions. On the other hand, T$_{C2}$, which separates the FM region and the one with a combination of AFM and FM, i.e., with canted spins, first rises up to $\sim$ 50 K for $x$ = 0.03 and then falls. An extrapolation suggests a visually quadratic variation of T$_{C2}$ with $x$. The canted spin structure can be anticipated to cease to exist beyond $\sim$ 6 $\%$ of Cr in CeFe$_2$. Such a dome-like canted-spin region has also been reported for Co substitutions in CeFe$_2$. \cite{Chaboy00} The data points for Ag and Au also have been shown in the phase diagram.

\begin{figure}[h]

\centering
  \centering
    \includegraphics[width=6in]{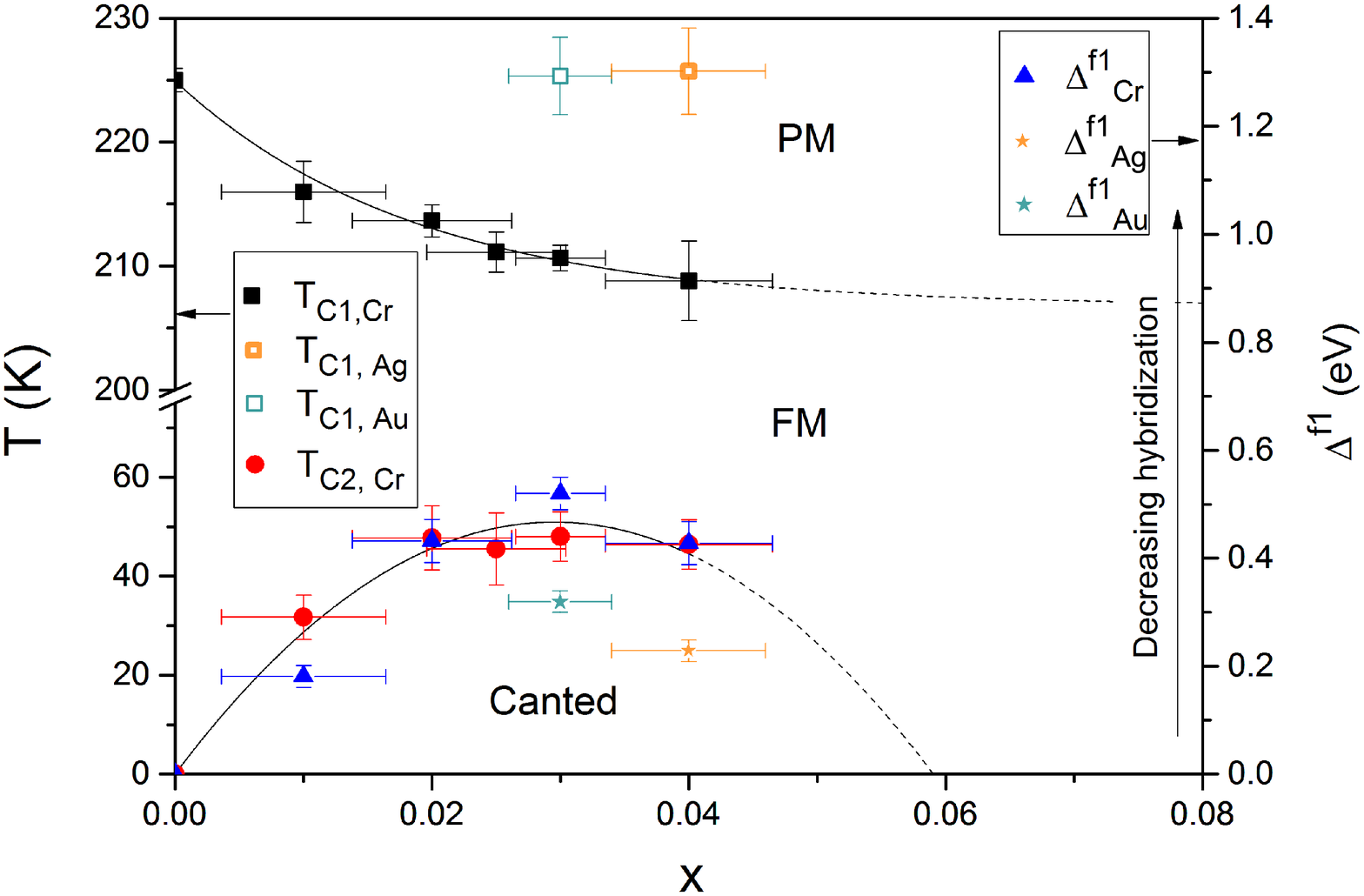}

\caption{Proposed T-$x$ phase diagram for the studied Ce(Fe$_{1-x}$M$_x$)$_2$ compounds. Additional subscripts Cr, Ag and Au have been used to differentiate between different impurities. Variation of $\Delta^{f1}$ with $x$ is also superimposed on the phase diagram.}

\label{figureeight}

\end{figure}

\subsection{X-ray Absorption Spectroscopy}

For XAS, we start with the Ce M$_{4,5}$ edges of Cr-series samples Ce(Fe$_{1-x}$Cr$_x$)$_2$, including the XAS of CeFe$_2$, because this is the series which shows variations in the magnetic properties. The spectra are shown in Fig. 9(a). It is known from theory and experiments \cite{Fuggle83XAS} that the Ce M$_{4,5}$ edge XAS spectra in Ce compounds are determined by transitions of type 3\emph{d}$^{10}$ 4\emph{f}$^n$ $\to$ 3\emph{d}$^9$ 4\emph{f}$^{n+1}$. The dominant transitions are specifically (i) from the initial state 3\emph{d}$^{10}$ 4\emph{f}$^0$, corresponding to Ce$^{4+}$, to 3\emph{d}$^{9}$ 4\emph{f}$^1$ final state, and (ii) from the initial state 3\emph{d}$^{10}$ 4\emph{f}$^1$, corresponding to Ce$^{3+}$, to 3\emph{d}$^{9}$ 4\emph{f}$^2$ final state. This happens for both the spin-orbit split components. In the spectra, the main peaks are due to the 3\emph{d}$^9$4\emph{f}$^2$ final-state multiplets, labeled as $f^2$, while the satellite structures $\sim$ 5 eV above the main peaks are due to the 3\emph{d}$^9$4\emph{f}$^1$ final states, labeled as $f^1$. Before presenting the spectra as in Fig. 9(a), any CeO$_2$ components \cite{Butorin96} were deconvoluted and subtracted. The `white lines' corresponding to the $f^2$ peaks of the M$_5$ components were then matched at the rising edge and their intensities normalized to 1. The presence of both the $f^1$ and $f^2$ peaks in the spectra corroborates the XPS results on the existence of mixed Ce valence and indicates a strong 4$f$-conduction electron ( i.e., $f-d$) hybridization in all the samples. \cite{Fuggle83XAS}

\begin{figure}[h]

\centering
\begin{subfigure}{.5\textwidth}
  \centering
    \includegraphics[width=5in]{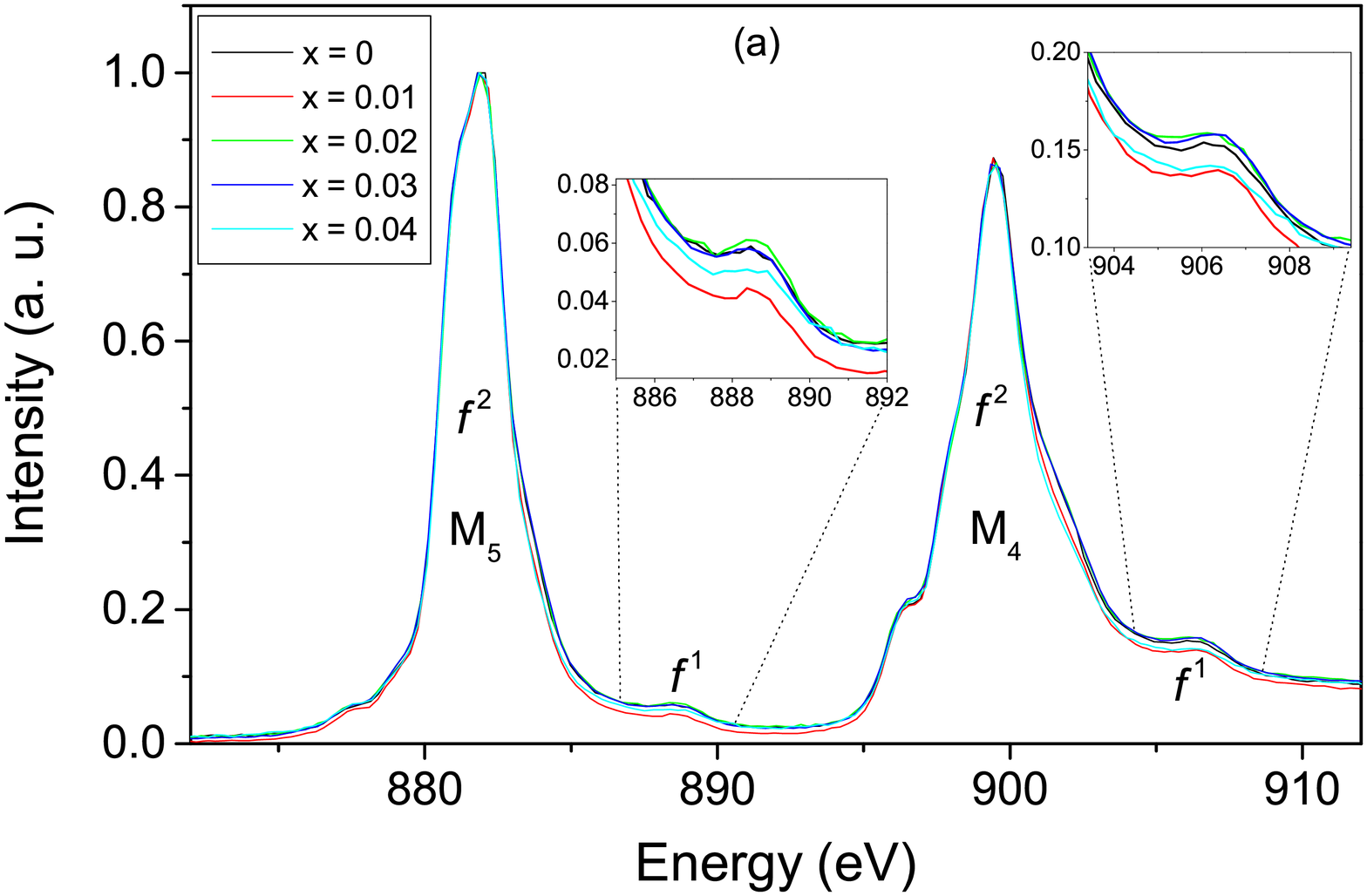}
\end{subfigure}%

\begin{subfigure}{.5\textwidth}
  \centering
    \includegraphics[width=5in]{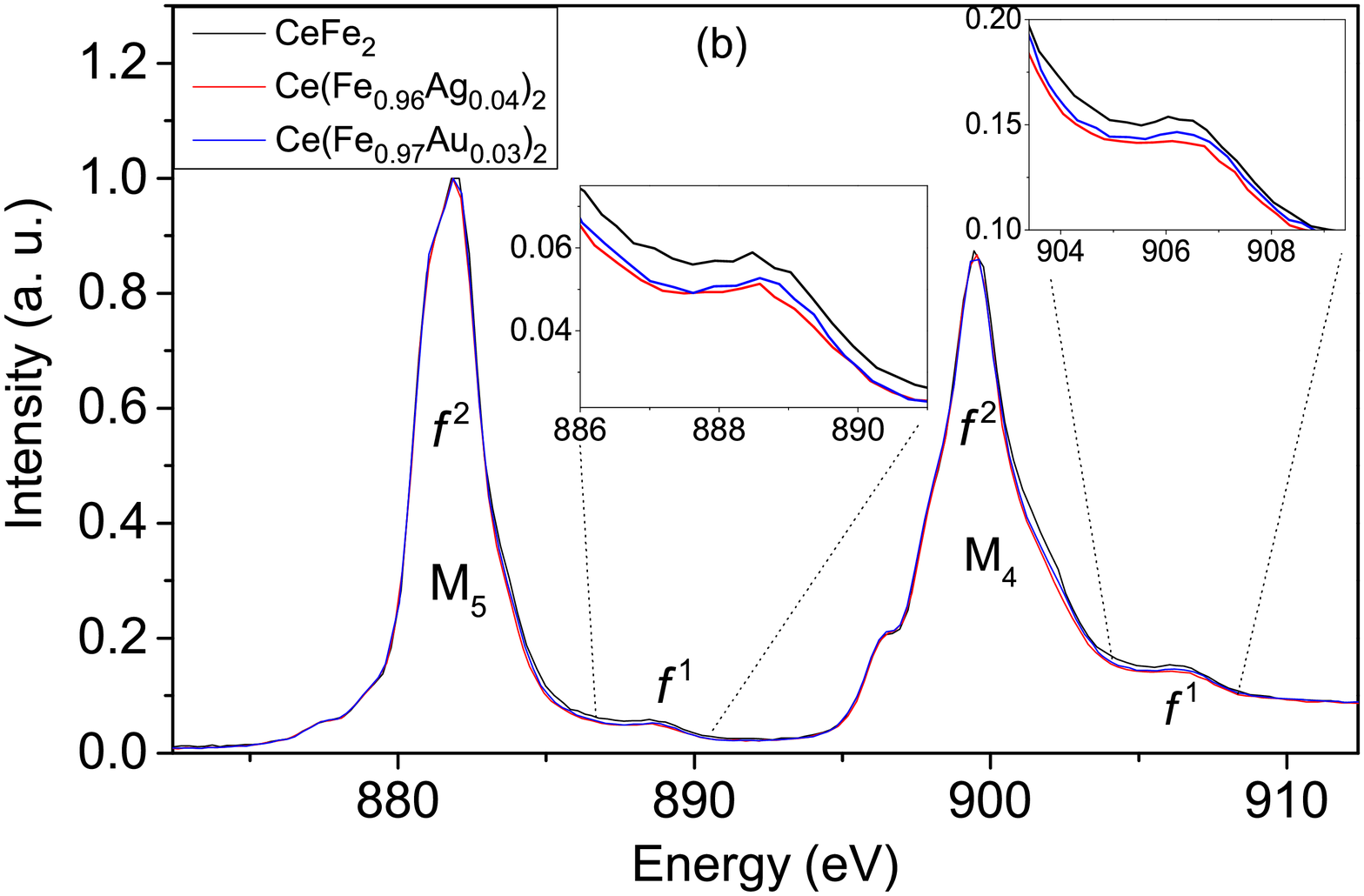}
\end{subfigure}%

\caption{Ce M$_{4,5}$ edge XAS spectra for (a) Cr series and (b) Ag and Au in comparison with CeFe$_2$. The insets show a magnified view of the spectra near the $f^1$ peaks.}

\label{figurenine}

\end{figure} 

In order to proceed further on the analyses of the XAS data, we refer to the theoretical work by Fuggle {\it et al.} \cite{Fuggle83XAS} cited above on intermetallic compounds of Ce. Based on an earlier XPS work, they assumed that the ground state wavefunction $\psi$ of Ce in such mixed valance compounds can be written as a mixture of 4$f^0$ (i.e., Ce$^{4+}$), 4$f^1$ (i.e., Ce$^{3+}$) and 4$f^2$ (i.e., Ce$^{2+}$) initial-state wavefuncitons $\phi^0, \phi^1$, and $\phi^2$, with weight factors $c_{(f^0)}$, $c_{(f^1)}$ and $c_{(f^2)}$, respectively, such that $\psi = (c_{(f^0)})^{1/2} \phi^0 + (c_{(f^1)})^{1/2} \phi^1 + (c_{(f^2)})^{1/2} \phi^2$. Then, using a modified Anderson impurity model, they calculated the 3$d$ $\to$ 4$f$ XAS spectra, which contain the two peaks $f^1$ and $f^2$ corresponding to the transitions 3$d^{10}$ 4$f^0$ $\to$ 3$d^{9}$ 4$f^1$ and 3$d^{10}$ 4$f^1$ $\to$ 3$d^{9}$ 4$f^2$, respectively, as also obtained in this work and mentioned above. The $f^1$ peak, according to their calculations, occurs at an energy 4 - 6 eV above the $f^2$ peak. The present results also maintain the $f^1$ - $f^2$ relative peak positions in the same range. They further showed that the relative intensity $I(f^1$)/[$I(f^1)+I(f^2$)] of the $f^1$ peak is equal to the weight $c_{(f^0)}$ of the Ce$^{4+}$ initial state, if there is no mixing of the $f$ levels and the conduction states, i.e., for zero $f-d$ hybridization width. If the hybridization is introduced and incremented, the relative intensity deviates from the equality with $c_{(f^0)}$, and for a large hybridization becomes even half of the $c_{(f^0)}$ value provided that $c_{(f^0)}$ remains unaltered. This way, in the present work where $I(f^2$) is made fixed, the variation in $I(f^1$) should give a measure of the variation in the hybridization strength with the Cr composition. Such $x$-dependent change in $I(f^1$) can indeed be visualized in Fig. 8 and its insets. However, its quantification is not straightforward. First, determination of a proper background for an $f^1$ peak is not unique, then there are possible multiplet effects and many-body tails as mentioned by Fuggle {\it et al.}, \cite{Fuggle83XAS} and finally any $x$-dpependent variation would be insignificant since the Cr concentrations are very small. A rough calculation using linear background, and taking the peak intensity as the area under the peak with the background subtracted, shows a very small ($\sim~10\%$) and unsystematic change in $I(f^1$) with $x$. This means that either essentially no information on the variation of the $f-d$ hybridization is availbale in the data, or there is no change in the hybridization with $x$. However, an interesting observation was made during this practice - the $f^1$ peak positions, which ought to be unaffected by the selection of the background to be subtracted, vary with $x$, although visually non-monotonically. In order to find out the cause of this shift, we reverted back to the theoretical results by Fuggle {\it et al.}. \cite{Fuggle83XAS} Their calculations do suggest that an incremental increase in $c_{(f^0)}$, i.e., in the $f-d$ hybridization, apart from increasing $I(f^1)$, is also associated with a systematic decrease in the relative peak position of $f^1$. Further, an examination of the plot between $I(f^1$)/[$I(f^1)+I(f^2$)] and $c_{(f^0)}$ in their work suggests that for nominal changes in $I(f^1$) as observed in the present results, an increase in the $f-d$ hybridization strength must correspond to an increase in $c_{(f^0)}$. Thus, if the theoretical model is valid, (an increase in) the $f^1$ - $f^2$ peak separation $\Delta^{f1}$, which is robust against the choice of the background, must also be a measure of (a decrease in) the $f-d$ hybridization strength. 

The plot of $\Delta^{f1}$ versus $x$ for Ce(Fe$_{1-x}$Cr$_x$)$_2$ series has also been shown superimposed on the magnetic phase diagram in Fig. 8, with $\Delta^{f1}$ on the right $y$-axis. As discussed above, the $\Delta^{f1}$ axis also represents a decreasing $f-d$ hybridization.  Interestingly, $\Delta^{f1}$ appears to follow exactly the same $x$ variation as does the T$_{C2}$. This indicates that the FM-AFM instability observable in CeFe$_2$ is to some extent directly determined by the change, in the present case the decrease, in the $f-d$ hybridization on impurity substitution. However, similar analyses performed on the XAS spectra of Ag and Au substituted CeFe$_2$ (Fig. 9(b)), i.e. in the cases where the impurity does not induce the second transition, the $\Delta^{f1}$ values, also plotted in Fig. 8, are finite. The decrease in the $f-d$ hybridization in all the three impurity cases is consistent with the previous computational results. \cite{Das16} Had the $f-d$ hybridization been the only mechanism driving the second transition, the two values would have been zero. Therefore, within the assumption that the theory reported by Fuggle {\it et al.} \cite{Fuggle83XAS} is true, it can be inferred from the present study that although the change in the $f-d$ hybridization is not a sufficient condition for impurity induced magnetic instability in CeFe$_2$, it is necessary, and has a definite proportionality with the second transition temperature.

\section{CONCLUSION}

We present in this work dc magnetization and Ce M$_{4,5}$ edge x-ray absorption spectroscopy studies of Ce(Fe$_{1-x}$M$_x$)$_2$ compounds, with M = Cr, Ag and Au, to seek any direct relation between the $f-d$ hybridization and the sub-Curie temperature second phase transition, if any, associated with the instability of ferromagnetism in CeFe$_2$. Two supplementary measurements - x-ray diffraction and x-ray photoelectron spectroscopy - are also performed to monitor the quality of the samples, the latter of which suggests that the oxidation states of Ce remain unaltered on substitution of Fe by the impurities M. Whereas the second transition is observable for Cr impurities, Ag and Au show none. The phase below the second transition temperature is inferred to be antiferromagnetic and the transition is found to be of first-order nature. The temperature-composition magnetic phase diagram derived from the magnetization data consists of three regions separated by the Curie and the second transition temperatures. Both the transition temperatures are found to vary systematically, but differently, with composition. The absorption data are analyzed on the basis of an existing theoretical report, wherein CeFe$_2$ x-ray absorption spectra for different $f-d$ hybridization strengths have been calculated and each spectrum is shown to consists of two peaks - $f^1$ and $f^2$. A correlation between the relative peak position and the hybridization strength can be derived from the spectra. Qualitative hybridization strengths calculated this way for the compounds with Cr impurity, which exhibit the second transition in the present work, are shown to follow the same composition dependence as does the second transition temperature. It is inferred that although the change in the $f-d$ hybridization is not a sufficient condition for impurity induced magnetic instability in CeFe$_2$, it is necessary, and has a definite proportionality with the second transition temperature.

\section*{ACKNOWLEDGEMENTS}

This work has been supported partially by the University Grants Commission, India and partially by the project CSR-IC-BL-28/CRS-125-2014-15/1221 from the UGC-DAE Consortium for Scientific Research, Indore, India. We also acknowledge the supports provided by the DST-FIST funded XPS and PPMS facilities in the Department of Physics, IIT Kharagpur, and the Central Research Facility, IIT Kharagpur.


\newpage


\begin{thebibliography}{11}

\bibitem{Kennedy90} S. J. Kennedy and B. R. Coles, J. Phys.: Condens. Matter \textbf{2}, 1213 (1990).

\bibitem{Paolasini03} L. Paolasini, B. Ouladdiaf, N. Bernhoeft, J-P. Sanchez, P. Vulliet, G. H. Lander, and P. Canfield, Phys. Rev. Lett. \textbf{90}, 057201-1 (2003).

\bibitem{royprb} S. B. Roy and B. R. Coles,  Phys. Rev. B \textbf{39}, 9360 (1989).

\bibitem{Roy89} S. B. Roy and B. R. Coles, J. Phys.: Condens. Matter \textbf{1}, 419 (1989).

\bibitem{Chaboy00} J. Chaboy, C. Piquer, L. M. Garc{\'i}a, F. Bartolom\'e, H. Wada, H. Maruyama, and N. Kawamura, Phys. Rev. B \textbf{62}, 468 (2000).

\bibitem{Roy04} S. B. Roy, G. K. Perkins, M. K. Chattopadhyay, A. K. Nigam, K. J. S. Sokhey, P. Chaddah, A. D. Caplin, and L. F. Cohen, Phys. Rev. Lett. \textbf{92}, 147203 (2004).

\bibitem{Haldar08} A. Haldar, K.G. Suresh, and A. K. Nigam, Phys. Rev. B \textbf{78}, 144429 (2008).

\bibitem{Vershinin14} A. V. Vershinin, V. V. Serikov, N. M. Kleinerman, N. V. Mushnikov, E. G. Gerasimov,
V. S. Gaviko, and A. V. Proshkin, Phys. MET. Metallography (USSR) \textbf{115}, 1208 (2014).

\bibitem{Chatt03} M. K. Chattopadhyay, S. B. Roy, A. K. Nigam, K. J. S. Sokhey, and P. Chaddah, Phys. Rev. B \textbf{68}, 174404 (2003).

\bibitem{Manekar01} M. A. Manekar, S. Chaudhary, M. K. Chattopadhyay, K. J. Singh, S. B. Roy, and P. Chaddah, Phys. Rev. B \textbf{64}, 104416 (2001). 

\bibitem{Sokhey04} K. J. S. Sokhey, M. K. Chattopadhyay, S. B. Roy, A. K. Nigam, and P. Chaddah, Solid State Commun. \textbf{129}, 19 (2004).

\bibitem{Wang12} J. Wang, Y. Feng, R. Jaramillo, J. van Wezel, P. C. Canfield, and T. F. Rosenbaum, Phys. Rev. B \textbf{86}, 014422 (2012).

\bibitem{Das16} R. Das, G. P. Das, and S. K. Srivastava, J. Phys. D: Appl. Phys. \textbf{49}, 165004 (2016).

\bibitem{Fukuda01} H. Fukuda, H. Fujii, H. Kamura, Y. Hasegawa, T. Ekino, N. Kikugawa, T. Suzuki, and T. Fujita, Phys. Rev. B \textbf{63}, 054405 (2001).

\bibitem{Forsthuber90} M. Forsthuber, F. Lehner, G. Wiesinger, G. Hilscher, T. Huber, E. Gratz, and G. Wortmann, J. Magn. Magn. Mater. \textbf{90 \& 91}, 471 (1990).

\bibitem{Konishi00} T. Konishi, K. Morikawa, K. Kobayashi, T. Mizokawa, A. Fujimori, K. Mamiya, F. Iga, H. Kawanaka, Y. Nishihara, A. Delin, and O. Eriksson, Phys. Rev. B \textbf{62}, 14304 (2000).

\bibitem{Fuggle80} J. C. Fuggle, M. Campagna, Z. Zolnierek, R. Laesser and A. Platau, Phys. Rev. Lett. \textbf{45}, 1597 (1980).

\bibitem{Fuggle83} J. C. Fuggle, F. U. Hillebrecht, Z. Zo\l{}nierek, R. L{\"o}sser, C. Freiburg, O. Gunnarsson, and K. Sch{\"o}nhammer, Phys. Rev. B \textbf{27}, 7330 (1983).

\bibitem{Zhang94} X. Zhang and A. Naushad, J. Alloys Compd. \textbf{207-208}, 300 (1994).

\bibitem{Eriksson88} O. Eriksson, L. Nordstrom, M. S. S. Brooks, and B. Johansson, Phys. Rev. Lett. \textbf{60}, 2523 (1988).

\bibitem{Levin02} E. M. Levin, K. A. Gschneidner, Jr., and V. K. Pecharsky, Phys. Rev. B \textbf{65}, 214427 (2002).

\bibitem{Roy05} S. B. Roy, M. K. Chattopadhyay, P. Chaddah, and A. K. Nigam, Phys. Rev. B \textbf{71}, 174413 (2005).

\bibitem{Mahen02} R. Mahendiran, A. Maignan, S. Hebert, C. Martin, M. Hervieu, B. Raveau, J. F. Mitchell, and P. Schiffer, Phys. Rev. Lett. \textbf{89}, 286602 (2002).

  
\bibitem{Singh02} K. J. Singh, S. Chaudhary, M. K. Chattopadhyay, M. A. Manekar, S. B. Roy, and P. Chaddah, Phys. Rev. B \textbf{65}, 094419 (2002).

\bibitem{Bustingorry16} S. Bustingorry, F. Pomiro, G. Aurelio, and J. Curiale, Phys. Rev. B \textbf{93}, 224429 (2016).

\bibitem{Fuggle83XAS} J. C. Fuggle, F. U. Hillebrecht, J.-M. Esteva, R. C. Karnatak, O. Gunnarsson and K. Sch{\"o}nhammer, Phys. Rev. B \textbf{27}, 4637 (1983).

\bibitem{Butorin96} S. M. Butorin, D. C. Mancini, J.-H. Guo, N. Wassdahl, J. Nordgren, M. Nakazawa, S. Tanaka, T. Uozumi, A. Kotani, Y. Ma, K. E. Myano, B. A. Karlin, and D. K. Shuh, Phys. Rev. Lett. \textbf{77}, 574 (1996).


\end{thebibliography}
\end{document}